\documentclass[useAMS,usenatbib]{mn2e}
\usepackage{graphicx}
\usepackage{textcomp}
\usepackage{txfonts}

\def \OIII {[O~{\sc iii}]5007~\AA}
\def \NII {[N~{\sc ii}]6584~\AA}
\def \ha {H$\alphaup$}

\def \vhel{\ifmmode{~V_{{\rm HEL}}}\else{~$V_{{\rm HEL}}$}\fi}
\def\msun{\ifmmode{{\rm\ M}_\odot}\else{${\rm\ M}_\odot$}\fi}

\def\myr{\ifmmode{{\rm\ M}_\odot{\rm\ yr}^{-1}}
         \else{${\rm\ M}_\odot$ yr$^{-1}$}\fi}

\def\tena#1 #2 {\ifmmode{#1 \times 10^{#2}}\else{$#1 \times 10^{#2}$}\fi}
\def\kms{\ifmmode{~{\rm km\,s}^{-1}}\else{~km~s$^{-1}$}\fi}

\def\~{$\sim$}

\title[Spatio-kinematic modelling of Abell 65]{Spatio-kinematic modelling of Abell~65, a double-shelled planetary nebula with a binary central star\thanks{Based on observations made with European Southern Observatory Telescopes at the La Silla Paranal Observatory, under programme ID \mbox{081.D-0857}}}
\author[Huckvale, Prouse et al.]{L. Huckvale,$^{1,2}$\thanks{E-mail:
leo.huckvale@postgrad.manchester.ac.uk} B. Prouse,$^{1}$ D. Jones,$^{3}$ M. Lloyd,$^{1}$ D. Pollacco$^{4}$, J.A. L\'opez,$^{5}$
\newauthor
T.J. O'Brien,$^{1}$ L. Sabin$^{6}$ and N.M.H. Vaytet $^{7}$
\\
$^{1}$Jodrell Bank Centre for Astrophysics, University of Manchester, M13 9PL, UK\\
$^{2}$European Southern Observatory, Karl-Schwarzschild-Stra\ss{}e 2, 85748 Garching, Germany\\
$^{3}$European Southern Observatory, Alonso de C\'ordova 3107, Casilla 19001, Santiago, Chile\\
$^{4}$Department of Physics, University of Warwick, Coventry, CV4 7AL, UK\\
$^{5}$Instituto de Astronom\'ia, Universidad Nacional Aut\'onoma de M\'exico, Apartado Postal 877, 22800 Ensenada, B.C., M\'exico\\
$^{6}$Instituto de Astonom{\'i}a y Meteorolog{\'i}a, Departamento de F{\'i}sica, CUCEI, Universidad de Guadalajara, Av. Vallarta 2602, C.P. 44130, Guadalajara, Jal., Mexico\\
$^{7}$Ecole Normale Sup\'{e}rieure de Lyon, CRAL (UMR CNRS 5574), 69364 Lyon Cedex 07, France}

\begin{document}

\date{Accepted 2013 June 17.  Received 2013 May 30; in original form 2012 November 9}

\pagerange{\pageref{firstpage}--\pageref{lastpage}} \pubyear{2013}

\maketitle

\label{firstpage}

\begin{abstract}

We present the first detailed spatio-kinematical analysis and modelling of the planetary nebula Abell 65, which is known to host a post-common envelope, binary, central star system.  As such, this object is of great interest in studying the link between nebular morphology and central star binarity.

\OIII{} and \ha{}+\NII{} longslit spectra and imagery of Abell~65 were obtained with the Manchester \'Echelle Spectrometer on the \mbox{2.1-m} telescope at the San Pedro Mart\'ir Observatory (MES-SPM). Further \OIII{} longslit spectra were obtained with the Ultraviolet and Visual \'Echelle Spectrograph on the Very Large Telescope (VLT-UVES). These data were used to develop a spatio-kinematical model for the \OIII{} emission from Abell 65. A ``best-fit'' model was found by comparing synthetic spectra and images rendered from the model to the data. The model comprises an outer shell and an inner shell, with kinematical ages of 15000~$\pm$~5000~yr~kpc$^{-1}$ and 8000~$\pm$~3000~yr~kpc$^{-1}$, respectively. Both shells have peanut-shaped bipolar structures with symmetry axes at inclinations of $(55\pm10)\degr{}$ (to the line-of-sight) for the outer shell and $(68\pm10)\degr{}$ for the inner shell. The near-alignment between the nebular shells and the binary orbital inclination (of $(68\pm2)\degr{}$) is strongly indicative that the binary is responsible for shaping the nebula. Abell~65 is one of a growing number of planetary nebulae (seven to date, including Abell~65 itself) for which observations and modelling support the shaping influence of a central binary.

\end{abstract}


\section{Introduction}

The Binary Hypothesis proposes that bipolar planetary nebulae (PNe) are shaped by interactions between the nebular progenitor star and a companion star \citep{demarco09}. A PN is formed during the final stages of the lifetime of a low mass star, as the giant progenitor star drives off a slow, dense stellar wind which is swept up by a later fast wind from the emerging white dwarf \citep{kwok78}. If the earlier dense wind develops an equatorial density enhancement, which restricts the outflow of the fast wind, the nebula will take on a bipolar, axi-symmetric \textit{hourglass} or \textit{diabolo} shape \citep{kahn85}. Such an enhancement could be generated by the presence of a companion star in close orbit with the PN progenitor (see \citealt{nordhaus06} and references therein). In this scenario, the progenitor star fills its Roche lobe and transfers matter to the companion star. Thermal instabilities in the companion star cause matter to overflow its Roche lobe and the system becomes enshrouded in a common envelope (CE) of stellar material \citep{iben93}. The CE may be shaped into a density enhancement in the orbital plane of the binary by conservation of angular momentum as the stars are slowed by the material and spiral inward (the CE is always shed in the orbital plane; \citealp{nordhaus06}). The Binary Hypothesis predicts that the symmetry axis of a bipolar nebula should be parallel to the orbital axis of the central binary system i.e. perpendicular to the orbital plane. Of the PNe known to have binary central stars, there is a small set in which the orbital parameters of their central stars are known.

To date, six such PNe have been shown by spatio-kinematical modelling to have nebular symmetry axes aligned with the orbital axes of the central binary stars, consistent with the Binary Hypothesis: Abell 63, \citep{mitchell07b}, NGC 6337 \citep{garcia09, hillwig10}, Abell 41 \citep{jones10}, Sp 1 \citep{jones2012a}, HaTr 4 \citep{tyndall12} and NGC 6778 \citep{miszalski11,guerrero12}.

Abell 65 (\mbox{PN G017.3$-$21.9}, \mbox{$\alpha=19^h46^m34.2$}, \mbox{$\delta=-23\degr08'12.9''$ J2000}; herein ``A~65''), discovered by \citet{abell66}, was described by \citet{bond90} to lie between the ``elliptical'' and ``butterfly'' types in the classification scheme of \citet{balick87}, having a major axis of 134\arcsec{} and a minor axis of 74\arcsec{}. Deep \OIII{} imagery obtained by \citet{bond90} shows filamentary structure to the north-east of the bright nebula described by \citet{abell66}, which they suggest could be bubble penetration perpendicular to the orbital plane. Deeper \OIII{} imagery obtained by \citet{walsh96} shows the full extent of the nebula, with fainter emission extending as far to the south-west as the filamentary structure in the north-east; they suggest that this may be polar outflow from the bright nebula. \citet{walsh96} also note a dark lane in the inner nebula which may correspond to a central cavity or ``hollow'' viewed along the line-of-sight near the symmetry axis. \citet{hua98} present observations of A~65 in \ha{}, \NII{} and \OIII{}, suggesting that the faint outer nebula (in \OIII{}) is a double ring-like structure surrounding the bright inner nebula. The inner nebula shows two bright lobes in their \NII{} image.

\citet{bond90} report an orbital period of $\simeq 1$ day for the central star. This is confirmed in photometric observations and modelling by \citet{Sh09}, who also use spectroscopic data to show that the binary is a young precataclysmic variable, having undergone a common envelope evolution. They also derive an inclination of the orbital plane of $(68\pm2)$\degr{} (where 90\degr{} corresponds to an eclipsing orbit).

Accurate distance measurements to PNe are notoriously difficult to obtain. However, the distance to A~65 has been measured by \citet{frew08}, using an H$\alpha$ surface brightness - radius relationship, to be 1.7~kpc and by \citet{stang2008}, using the Shklovsky/Daub distance technique (recalibrated using PNe in the Magellanic clouds), to be 1.672~kpc. For this study we adopt a distance of 1.7~kpc.

In this paper we present imagery and longslit spectroscopy of A~65, from which we derive a spatio-kinematical model of the nebula, in order to investigate the relationship between the nebular morphology and its central star.

\begin{figure*}
\centering
\includegraphics[trim=0 0 60 250, clip, width=\textwidth]{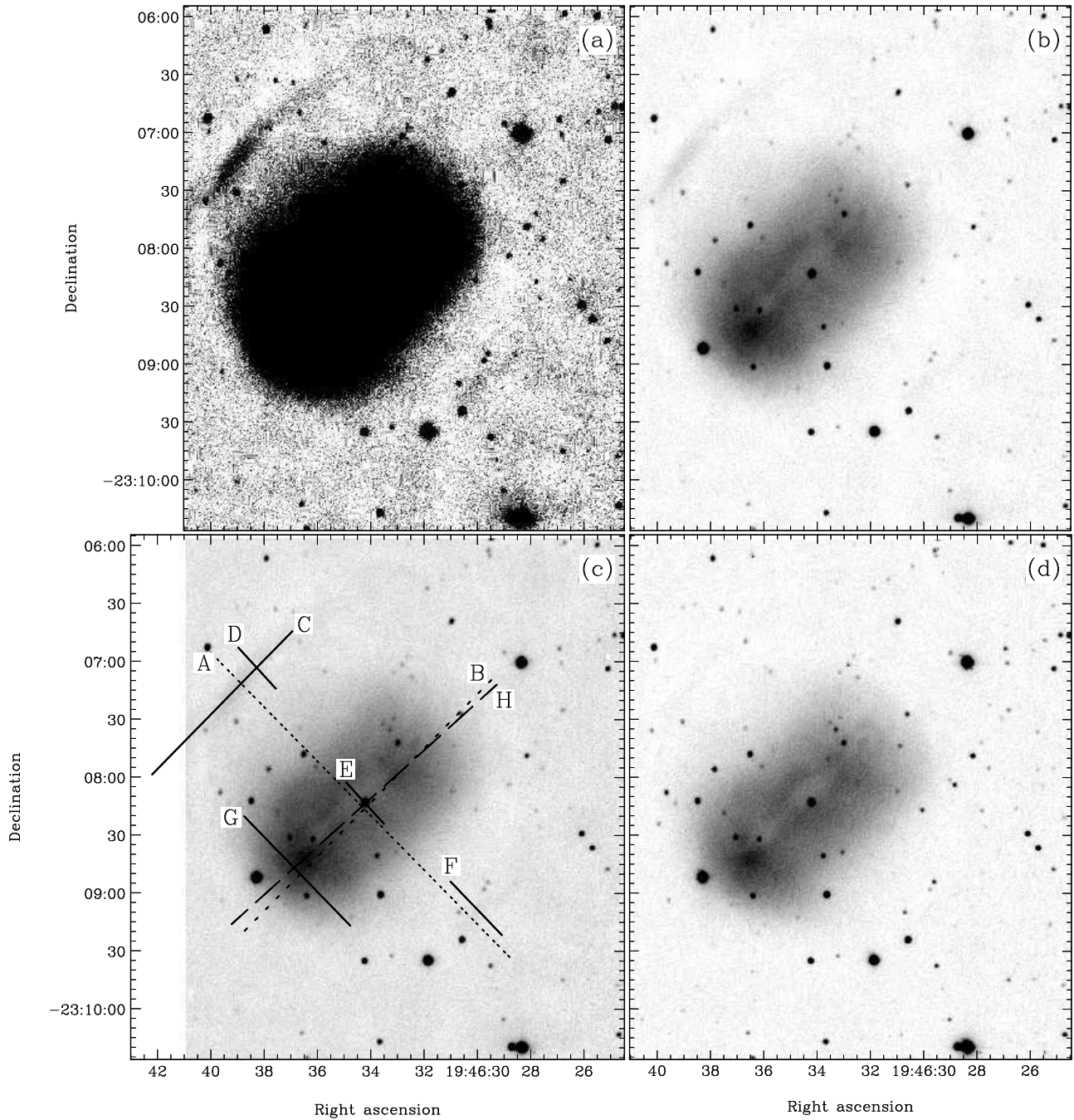}
\caption{Deep \OIII{} (a, b) and \ha{}+\NII{} (d) imagery of A~65 obtained with MES-SPM. (a) shows the same data as (b) with contrast adjusted to show the full extent of the nebula. The images were obtained within 2 hours of twilight, with a second quarter Moon 42\degr{} away, through an airmass of 1.72. The sky background across the images had a noticeable slope, which was fitted and subtracted from the images shown here. (c) shows the same data as (b) with the positions and sizes of the slits used to obtain the spectra.  Dotted lines correspond to MES-SPM \OIII{} observations (slits (A, B)), solid lines correspond to VLT-UVES \OIII{} observations (slits C-G) and the long-dashed line corresponds to MES-SPM observations in \ha{} and \NII{} (slit H). In each case, the label corresponds to the top of the slit and the top of the PV-arrays shown in Figures \ref{fig:spectraMESOIII}, \ref{fig:spectraUVESOIII} and \ref{fig:spectraMESHaNII}. The blank region for right ascension $>$19:46:41 is included to show the full length of slit C, but contains no image data.}
\label{fig:images}
\end{figure*}

\section{Observations and Results}
\subsection{\OIII{} and \ha{}+\NII{} imagery}

Narrowband images of A~65 were acquired with the Manchester \'Echelle Spectrometer (MES-SPM; \citealp{meaburn03}) on the \mbox{2.1-m} telescope at San Pedro Mart\'ir on 2008 June 17, two each in \OIII{} (using a 50\AA{} bandpass filter) and \ha{}+\NII{} (90\AA{} bandpass), with each exposure being 30 minutes. These were obtained using the SITe3 $1024\times 1024$ pixel CCD with $2\times 2$ binning, giving a pixel scale of 0.62\arcsec{} pixel$^{-1}$. Seeing was $\simeq 1.8\arcsec{}$ during these observations. The data were bias-corrected, flat-fielded, cleaned of cosmic rays and co-added using standard {\sc starlink} routines\footnote{See: http://starlink.jach.hawaii.edu/docs/sun86.htx/sun86.html}. The resulting images are shown in Figure \ref{fig:images}.

The \OIII{} imagery in Figure~\ref{fig:images}(a, b) shows the same bright inner nebula originally observed by \citet{abell66}. The inner nebula appears nearly rectangular with a major axis aligned south-east to north-west and measures $152\arcsec{} \times 86\arcsec{}$. There is a bar of faint emission at about 90\arcsec{} to the north-east and, in Figure~\ref{fig:images}(a), a counterpart at about 82\arcsec{} to the south-west of the central star. (This faint emission will later be referred to as the \textit{outer} nebula.) The full extent of the nebula can be seen in Figure~\ref{fig:images}(a), which shows the same data as Figure~\ref{fig:images}(b) at high contrast. The \ha{}+\NII{} imagery, in Figure~\ref{fig:images}(d), shows the bright inner nebula with similar dimensions to the \OIII{} image. The outer nebula is not detected in \ha{}+\NII{}.

\subsection{Longslit \OIII{} spectroscopy}

\begin{figure*}
\centering
\includegraphics{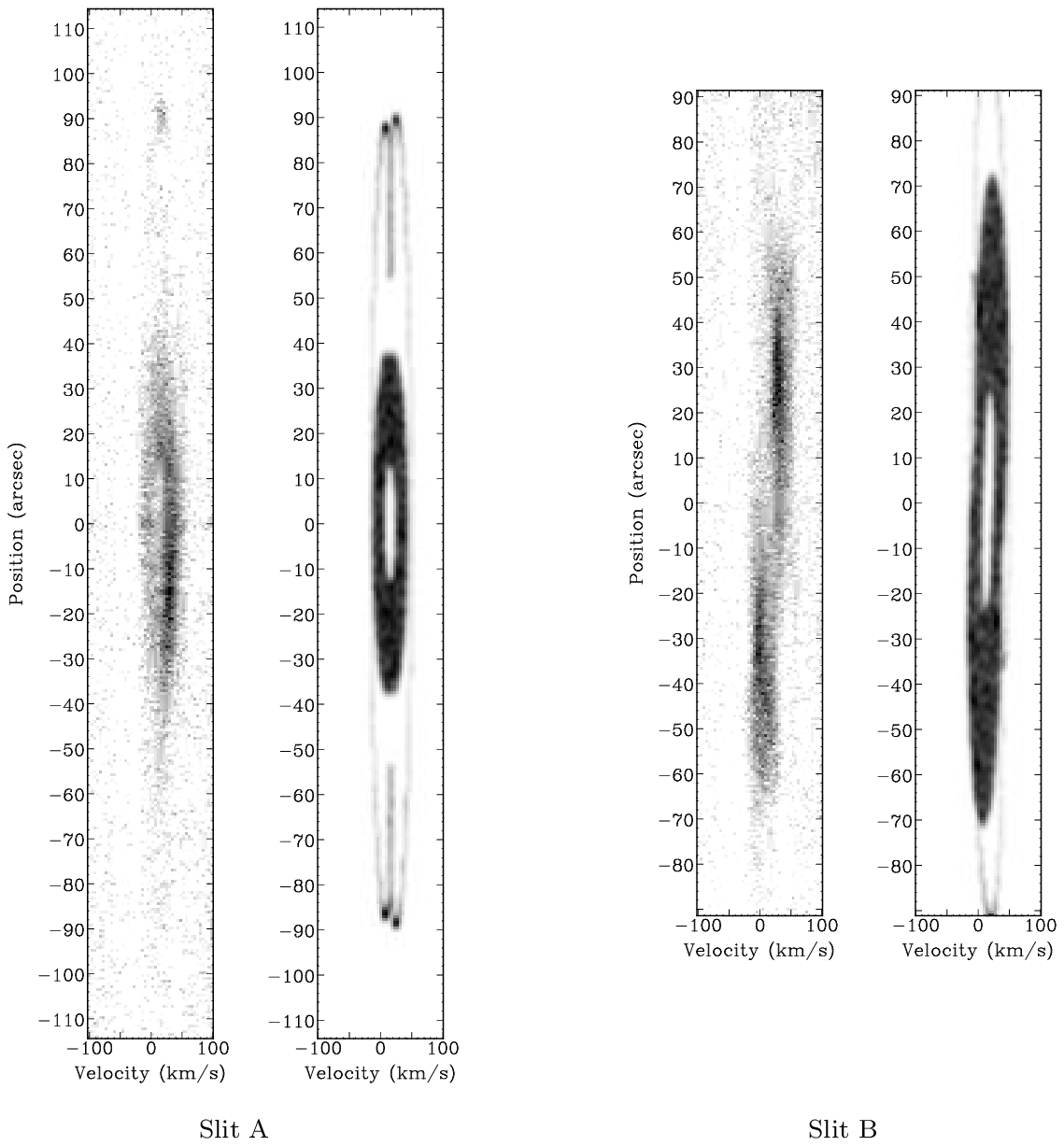}
\caption{Longslit \OIII{} spectra of A~65 acquired with MES-SPM for slits A and B (left in each case), and their corresponding synthetic spectra rendered from the spatio-kinematical model (right in each case). The positions of these slits are shown in Figure \ref{fig:images}(c). The spectra have been calibrated to show heliocentric velocity.}
\label{fig:spectraMESOIII}
\end{figure*}

\begin{figure*}
\centering
\includegraphics{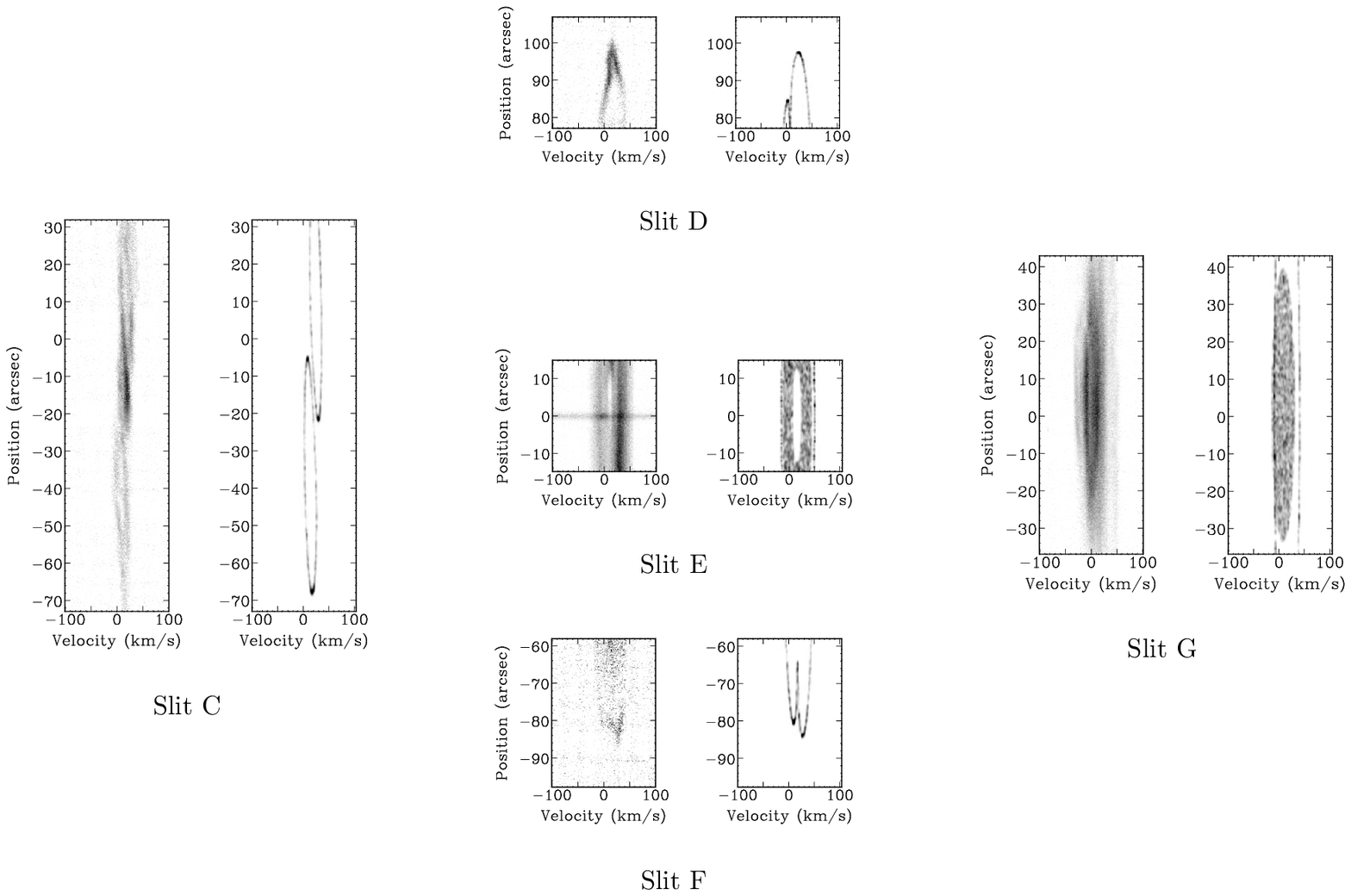}
\caption{Longslit \OIII{} spectra of A~65 acquired with VLT-UVES for slits C-G (left in each case) and their corresponding synthetic spectra rendered from the spatio-kinematical model (right in each case). The positions of these slits are shown in Figure \ref{fig:images}(c). Slits D-F are parallel and approximately aligned, and are laid out in this figure to be in correct relative positions for useful comparison with slit A in Figure \ref{fig:spectraMESOIII}. The spectra have been calibrated to show heliocentric velocity.}
\label{fig:spectraUVESOIII}
\end{figure*}

\begin{figure*}
\centering
\includegraphics[clip]{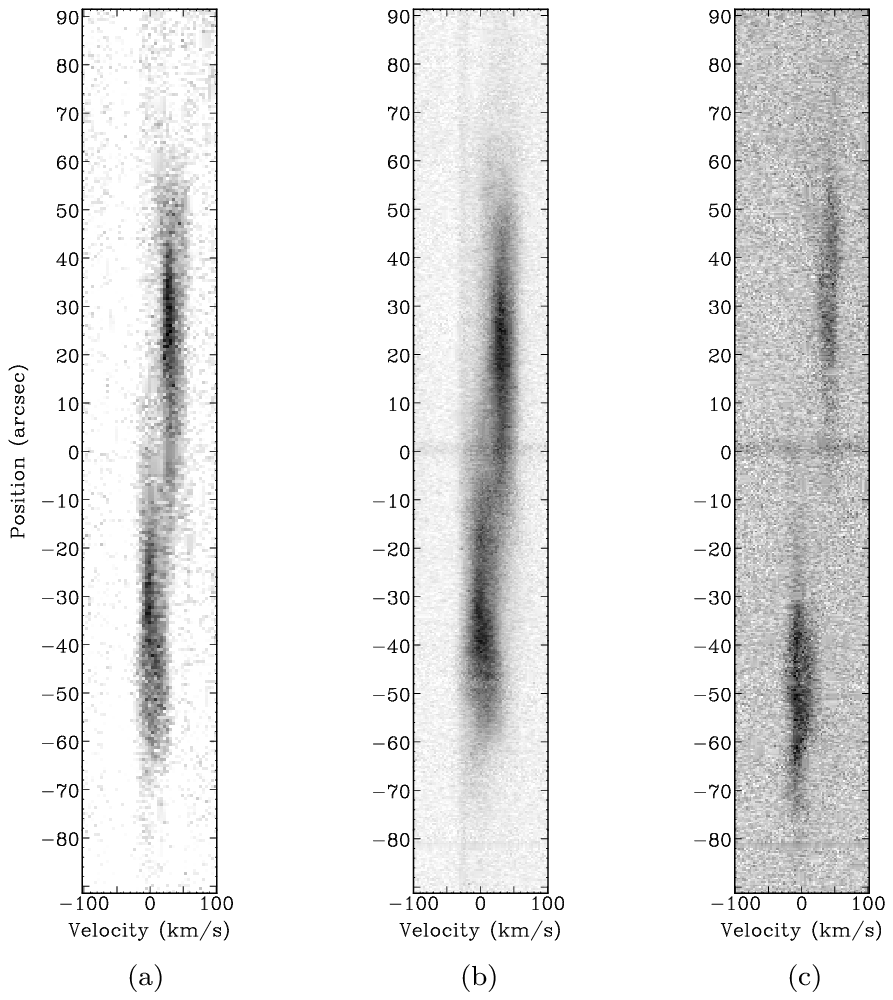}
\caption{Longslit spectra acquired with MES-SPM in (a) \OIII{}, (b) \ha{} and (c) \NII{}. The \OIII{} spectrum is that of slit B (Figure \ref{fig:images}(c)), while the other two spectra were acquired with slit H, with approximately the same orientation as slit B. The spectra have been calibrated to show heliocentric velocity.}
\label{fig:spectraMESHaNII}
\end{figure*}

Spatially resolved, longslit emission line spectra of A~65 were also obtained with MES-SPM. Observations were made on 2008 June 18 using a narrowband filter centered on \OIII{} to isolate the 114th \'echelle order. Seeing was $\simeq 2\arcsec{}$. A slit 30~mm long ($\equiv$ 5\arcmin{} on the sky) and 150~$\muup$m wide ($\equiv$~2\arcsec{}) gave a spectral resolution of $R\simeq 27,000$, equivalent to 11\kms{}. The SITe3 CCD with $2\times{}2$ binning gave a pixel scale of 0.62\arcsec{} in the spatial axis and 4.79\kms{}~pixel$^{-1}$ in the spectral axis. The exposure time for these observations was 1800 seconds. The slit positions are shown in Figure~\ref{fig:images}(c) as A and B, covering the minor and major axes (respectively) of the bright nebula.

Further longslit spectra, in \OIII{}, were obtained using VLT-UVES \citep{dekker00,messenger06}, in single-order mode, on 2008 July 11 and 12. The seeing for these observations never exceeded 1\arcsec{}. These observations used the EEV CCD in the ``red'' (visible to near-infrared) arm of the spectrograph, and had a spatial pixel scale of 0.16\arcsec{} pixel$^{-1}$ and a velocity pixel scale of 1.2\kms{}~px$^{-1}$. With a slit width of 0.6\arcsec{} VLT-UVES has a resolution of $R\simeq 70,000$ in the red arm, equivalent to 4\kms{} \citep{dekker00}. The exposure time for these observations was 1200 seconds. The slit on VLT-UVES is only 30\arcsec{} long on the sky, so longer effective slit-positions were reconstructed from several partially overlapping observations. Five slit-positions C-G were reconstructed from a total of eleven VLT-UVES observations, as shown in Figure~\ref{fig:images}(c).

Data reduction for all spectra was performed using {\sc starlink} software. The spectra were bias-corrected and cleaned of cosmic rays, then wavelength-calibrated against a ThAr emission-lamp to an absolute accuracy of $\pm~1.5$\kms{}. Finally the data were rescaled to a linear velocity scale, relative to the rest wavelength of [O~{\sc iii}] 5006.843~\AA{}, and corrected to show heliocentric velocity, $V_{hel}$. The spatial axis for each spectrum was offset such that 0\arcsec{} corresponds to a line drawn perpendicular to the slit, through the central star. The reduced spectra from MES-SPM and VLT-UVES are shown in Figures \ref{fig:spectraMESOIII} and \ref{fig:spectraUVESOIII}, respectively.

The \OIII{} spectrum from slit A (see Figure \ref{fig:spectraMESOIII}), aligned along the minor axis of the bright inner nebula, shows faint emission at the north-east end of the slit at about 90\arcsec{} from the centre of the nebula. This coincides with the position of the faint emission from the outer nebula seen in
MES-SPM \OIII{} imagery (Figure \ref{fig:images}). Slit A also shows emission from the bright inner nebula as a velocity ellipse extending from about -50\arcsec{} to about 40\arcsec{} in the spatial axis, and from about -20\kms{} to about 50\kms{} in the spectral axis. The spectrum from slit B, aligned along the major axis of the nebula shows two bright components, one redshifted towards the north-west end of the slit and one blueshifted towards the south-east end.

Slit C covers the faint emission from the outer nebula, at about 90\arcsec{} to the north-east of the central star. The \OIII{} spectrum from this slit shows two closed velocity ellipses, redshifted and blueshifted to the north-west and south-east ends of the slit, respectively, in the same sense as the two bright components of the inner nebula seen in the spectrum from slit B.
Slits D-F are parallel and near to the line of slit A, providing more detailed information on the outer nebula to the north-east (D), the centre of the inner nebula (E) and the outer nebula to the south-west (F). Slit D and slit F each show a partial velocity ellipse with a ``kinked'' shoulder. The spectrum from slit E shows two velocity components running the length of the slit (likely the centre of the larger velocity ellipse seen in slit A). Each of these components is $\simeq 20\kms{}$ in full-width at half-maximum and their centres are separated by $\simeq 30\kms{}$.
Slit G is parallel to the minor axis of the inner nebula and offset from the central star by about 50\arcsec{}, and shows a bright velocity ellipse $\simeq 30\kms{}$ wide, with an unfilled central region $\simeq 10\kms{}$ wide.

\subsection{Longslit \ha{} and \NII{} spectroscopy}

Images obtained by \citet{hua98} show clear differences between the \NII{} and \OIII{} emission regions. In order to compare the \NII{} (and \ha{}) emission with the longslit \OIII{} observations, a later observation was made with MES-SPM on 2010 September 24, this time using a narrowband (90~\AA) filter, centered on the 87th \'echelle order, to obtain \ha{} and \NII{} longslit spectra in the same frame. Seeing was $\simeq 2\arcsec{}$. This observation used a TH2K $2048\times 2048$ pixel CCD in $2\times 2$ binning, with a spatial pixel scale of 0.36\arcsec{} pixel$^{-1}$ and a spectral pixel scale of 2.1\kms{} pixel$^{-1}$. The slit width (150~$\muup$m~$\equiv$~2\arcsec{}) and resolution ($R\simeq 27000~\equiv~11\kms$) were the same as in the \OIII{} MES-SPM observations. Data reduction was carried out as before, using the appropriate rest wavelengths of H$\alphaup$ 6562.81~\AA{} and [N~{\sc ii}] 6583.45~\AA{} to rescale the spectral axis to heliocentric velocity, $V_{hel}$. The exposure time for these observations was 1800 seconds. The position of this slit, H, is shown in Figure~\ref{fig:images}(c) and its \ha{} and \NII{} spectra are shown alongside the \OIII{} spectrum, from slit B, in Figure~\ref{fig:spectraMESHaNII}.

In \ha{}, slit H shows a pair of velocity ellipses (Figure \ref{fig:spectraMESHaNII}(b)) in similar orientation and position to those in the \OIII{} spectrum from slit B, although the central region between the ellipses appears more filled in \ha{}. The \NII{} spectrum (\ref{fig:spectraMESHaNII}(c)) shows a markedly different distribution to both \ha{} and \OIII{}, with much less emission in the central region.

\section{Discussion}
\subsection{Nebular structure}

A~65 was previously thought (based solely on imagery) to have a symmetry axis aligned northeast-southwest with the inner nebula being an equatorial ring, viewed almost edge-on, and the outer nebula being a polar outflow \citep{bond90, walsh96}.

Slit A shows two bright components, separated by $\simeq 20\kms{}$ at 0\arcsec{} spatial offset, forming a closed velocity ellipse corresponding to the inner nebula. This suggests that the structure of the inner nebula is a closed shell along the northeast-southwest axis and therefore cannot be a ring with polar outflow along a northwest-southeast axis, as previously proposed.

Slit B, though showing little or no emission from the fainter nebula, shows a red-shifted component towards the top of the slit (north-west) and a corresponding blue-shifted component towards the bottom (south-east). The peak brightnesses of the lobes are separated by 63\arcsec{} and 27\kms{}. The components are suggestive of two lobes of a bipolar structure with a symmetry axis roughly parallel to the slit orientation, but at some inclination to the line-of-sight. This orientation is consistent with that suggested by slit A.

Slit C provides key information on the structure of the outer nebula. A double velocity ellipse indicates a bipolar structure, as in the case of the inner nebula, in contrast with the ring structure suggested by \citet{hua98}. The orientation of slit C is roughly parallel to the symmetry axis of the bright inner nebula, just as for slit B. The faint emission regions of the outer nebula to the north-east and south-west are almost equidistant from the centre of the inner nebula. This suggests that the nebula is a double-shelled structure, the outer shell being concentric with the inner shell.

Slits D, E and F confirm, at higher resolution and sensitivity, the structures observed in slit A. Slit D makes clear the kinked structure at the top of slit A. Slit E shows the central components more clearly. Slit F detects structure from the outer nebula not seen in slit A. Slit G appears to show an open velocity ellipse, similar to the inner nebula detected in slit A.

The \ha{} emission detected with slit H, in Figure \ref{fig:spectraMESHaNII} follows the \OIII{} emission, as would be expected from ionisation stratification. The \NII{} emission detected with slit H may correspond to the lobe tips of the inner nebula. This interpretation is also consistent with the \NII{} spectrum obtained with a north-south slit, included in the SPM Kinematic Catalogue of Planetary Nebulae\footnote{http://kincatpn.astrosen.unam.mx}\citep{lopez2012}.

\subsection{Spatio-kinematical modelling of Abell 65}

\begin{figure*}
\centering
\includegraphics[scale=0.5]{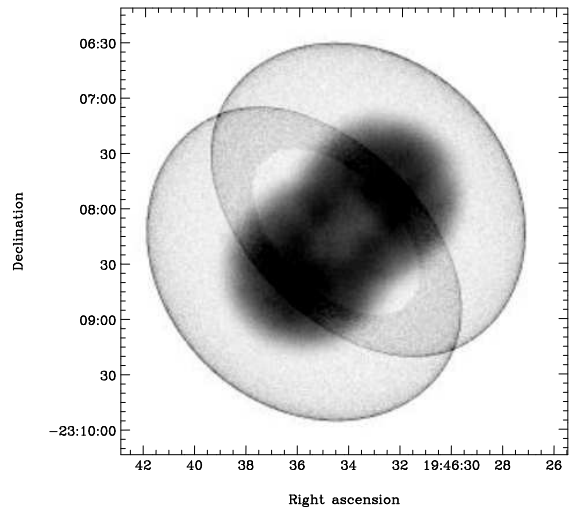}
\caption{Synthetic sky-image rendered from the spatio-kinemetical model of Abell 65. Shown at the same inclination as in the images in Figure \ref{fig:images}}
\label{fig:model}
\end{figure*}

A three-dimensional spatio-kinematical model of A~65 was developed from the images and spectra of the \OIII{} emission using {\sc shape} \citep{steffen11}. The inner and outer shells were modelled separately, given the distinct inner and outer emission regions in the images and spectra. For each shell radial expansion velocity was assumed to be proportional to the distance from the centre. Model parameters -- determining size, shape, expansion velocity, position angle and inclination -- were manually varied over a wide range. The best-fit model for each shell was determined by comparing, by eye, the data synthesised from the model to the observed \OIII{} image and \OIII{} longslit spectra. This is the same method which has been used to model Abell~41 \citep{jones10}, Sp~1 \citep{jones2012a} and HaTr~4 \citep{tyndall12}. The synthetic spectra are presented alongside the observed spectra in Figures \ref{fig:spectraMESOIII} and \ref{fig:spectraUVESOIII}. In each case, the synthetic spectra are rendered at the same spatial and spectral resolution as the observed spectra. It is the relation of the overall shape of the emission regions to parameters, particularly the orbital inclination, of the central binary star which is of interest, so no attempt was made to model the luminosity variations across the nebula. This would otherwise require a full radiative transfer treatment which is beyond the aims and scope of this work. Uncertainties in values derived from the model were determined by assessing the range of values over which the synthesised data were comparable to the observed data. The best-fit models of both shells are combined in a synthesised image in Figure \ref{fig:model}.

The best-fit model for the outer nebula was found to be a thin-shelled peanut shape with a symmetry axis aligned south-east to north-west at an inclination (to the line-of-sight) of $(55\pm 10)\degr{}$. The lobes of this model shell had to be large enough to produce the two closed velocity ellipses seen in slit C. However, there is insufficient evidence to determine whether the outer nebula is fully closed, as in the model, or open-ended beyond the region covered by the slit. Slit B, which would cover the full length of the outer nebula, does not show any emission other than that from the bright inner nebula, which does appear to be closed. However, this lack of emission from the outer nebula in slit B may be partly due to the lesser sensitivity of MES compared to UVES.

The best-fit model for the inner nebula was found to be a thick-shelled peanut shape, analogous to the outer nebula, with a symmetry axis aligned south-east to north-west at an inclination (to the line-of-sight) of $(68\pm 10)\degr{}$. Slits A and B together strongly suggest a bipolar shape with a symmetry axis approximately parallel to the position angle of slit B, the northwest lobe being inclined away from the observer. (Models with other orientations were attempted, such as those orthogonal orientations suggested by previous authors such as \citealt{bond90,walsh96}, but none were successful.) However, finding a model shape (thickness and extent) which could satisfactorily reproduce the observed ``hollow'' structure proved difficult. The hollow cavity in the model is limited by the extent of the shell in the observed image and in the PV-array from slit B, but a model which is consistent with these data fails to reproduce the velocity ellipse observed in slit G. The data from slit G suggest that in fact the central cavity extends further along the symmetry axis. Most likely, either a complex variation of shell thickness or deviation from axi-symmetry would be required to reconcile these issues.  Here we are concerned primarily with broad-scale structure and inclination, and so we have not pursued the detailed shape of the nebula further, rather accepting the simplest best-fitting model with replicates the main features of the observed images and spectroscopy. 

The parameters of the outer and inner nebula models are given in Table \ref{tab:dimensions}, for a unit distance 1~kpc and the measured distance of 1.7~kpc \citep{frew08,stang2008}.

\begin{table*}
\caption{The dimensions and kinematical ages of the outer and inner nebula shells per kpc of distance to the nebula.}
\label{tab:dimensions}
\begin{tabular}{lcccc}
							&	\multicolumn{2}{| c |}{Outer nebula}	&	\multicolumn{2}{| c |}{Inner nebula}	\\
							&	per kpc		&	1.7 kpc					&	per kpc			&	1.7 kpc				\\
Symmetry axis length		&	0.93 pc		&	1.6 pc					&	0.72 pc			&	1.2 pc				\\
Minimum waist diameter		&	0.57 pc		&	1.0 pc					&	0.36 pc			&	0.6 pc				\\
Maximum lobe diameter		&	0.91 pc		&	1.5 pc					&	0.42 pc			&	0.7 pc				\\
Waist expansion velocity	&	\multicolumn{2}{| c |}{17 \kms{}}		&	\multicolumn{2}{| c |}{22 \kms{}} 		\\
Polar expansion velocity	&	\multicolumn{2}{| c |}{31 \kms{}}		&	\multicolumn{2}{| c |}{44 \kms{}} 		\\
Kinematical age		&	$15000\pm5000$~yr	&	$25500\pm8500$~yr	&	$8000\pm3000$~yr	&	$13600\pm5100$~yr	\\
\end{tabular}
\end{table*}

\subsection{Systemic velocity and kinematical ages}

The systemic heliocentric radial velocity of A~65 was determined from the velocity shift between the observed spectra and the synthetic spectra rendered from the best-fit model, the model being effectively at rest. This was found to be approximately $13\kms{}$, consistent with the published value, $(13\pm4)\kms{}$, of \citet{meatheringham88}.

The best-fit model of the inner shell has an expansion velocity ranging from 22-44\kms{} from the waist to the poles. The kinematical age for the inner shell derived from this model, is $(8000\pm3000)$~yr~kpc$^{-1}$, giving an age of $(13600\pm5100)$~yr at the measured distance of 1.7~kpc. The best-fit model of the outer shell has an expansion velocity ranging from 17-31\kms{} from the waist to the poles. The kinematical age for the outer shell, derived from this model, is $(15000\pm5000)$~yr~kpc$^{-1}$, giving an age of $(25500\pm8500)$~yr at 1.7~kpc (the kinematical ages for both shells are also shown in Table \ref{tab:dimensions}). The outer shell appears older than the inner shell, potentially because it has been slowed by the ISM, although it is worth noting the large errors. These uncertainties, as mentioned in Section 3.2, were determined from the range over which the model gave an acceptable fit to the data.

\subsection{Context}

A~65 is the first PN with a known binary central star to show strong \textit{spectroscopic} evidence of a double-shelled structure. \citet{miszalski2009} identified a further four ``multiple-shelled'' PNe based on their deep imaging survey (PHR~J1804--2913, Pe~1-9, K~6-34 and Hf~2-2). However, without the addition of spatio-kinematical observations such as those presented in this paper it is difficult to fully characterise the nature of these outer shells\footnote{This need for detailed spatio-kinematical studies is remarked upon in the paper of \citet{miszalski2009}, indeed with particular reference to the case of A~65.}. Based on imagery alone it is difficult to distinguish between thin shell structures (such as the outer shell of A~65) and halo-like structures (for example, \citealp{corradi2003,corradi2004}). This is an important distinction as haloes are generally understood to be the remnant of the mass lost on the AGB which is being swept up into the nebular shell, whereas the mechanism by which multiple, distinct thin shells form is still unclear. It has been hypothesised that only a central binary could be responsible for such rapid changes in mass-loss as those required by the GISW theory to form more than one shell (\citealp{chu1989} and references therein). However, this may also be possible through a combination of photo-ionisation and wind interaction \citep{schoenberner1997}. Although the kinematical ages derived for the two shells have large uncertainties, they appear not to be coeval, the outer shell having nearly twice the kinematical age of the inner shell. This may support a binary-driven origin for the double shell structure, although it does not rule out other scenarios.

Assuming that the two shells identified here are indeed separate, non-coeval shells, this necessitates rapid evolution of mass-loss in the system as each shell requires a phase of enhanced mass-loss, and, based on their ages, they formed only $\sim$12000 years apart.  It is intriguing to consider that this is approximately the duration of an AGB thermal pulses, however there is no physical reason to two shells during a single pulse - rather this would require two subsequent pulses and this can be all but ruled out given that the interpulse period is some $~10$ times greater \citep{paxton11}. An alternative explanation for such a rapid change in mass-loss might be that the central star experienced a born-again phase after the ejection of the outer shell \citep{iben96}.  This would seem unlikely as the central star system shows no deficiency in Hydrogen even at light minimum where the majority of the flux originates from the primary \citep{Sh09}.

The simplest, and therefore most plausible, origin for the double-shelled structure is that they are actually formed by two jet-like ejections, or periods of collimated fast wind (CFW; \citealp{soker01}). These could then be expected to have evolved on binary time-scales and therefore account for the rapid changes of mass-loss. The only problem with this scenario is that it necessitates high mass-loss rates between the two ejection events (otherwise there will be insufficient material to form the inner shell), but an accepted requirement of the CFW models is that the slow, dense and collimated fast winds are blown (almost) simultaneously \citep{soker00,frank98}. Hence, the real test of this hypothesis is how the CFW was initiated (forming the outer shell), then stopped for some 10000 years and then re-initiated in order to form the inner shell. The source of this is unclear but almost certainly related to the evolution of the binary central star, which merits a much more detailed analysis.

 Interestingly, further support for the CFW formation hypothesis can be found in the distribution of \NII{} in A~65, which seems to be confined to the lobe tips of the inner shell (Fig. \ref{fig:spectraMESHaNII}(c) here, and Fig. 2(b) of \citealp{hua98}).  This can then be explained as the ``remnant'' shock ionisation from the ejection event that produced the inner shell.

\section{Conclusions}

Detailed spatio-kinematical modelling, based on deep imagery and high resolution spatially resolved spectroscopy, has shown that the bright inner nebula of A~65 has a bipolar structure which is oriented northwest-southeast. This is roughly perpendicular, in position angle, to that which was previously believed (prior observations of the nebula had only included image data). Our analysis has also shown that the nebula comprises two shells, similar in shape and inclination to each other. The faint emission observed to the north-east and south-west of the inner nebula has been shown to be a second bipolar shell in similar orientation, rather than a polar outflow or bubble penetration, as previously thought.

A~65 is the first PN with a known binary central star to have a proven double-shelled structure.
The inclinations of the symmetry axes of the inner and outer shells ($(68\pm 10)\degr{}$ and $(55\pm 10)\degr{}$ respectively) agree reasonably well with the orbital inclination of the central binary star ($(68\pm 2)\degr{}$), supporting the hypothesis that A~65 has been shaped by its binary central star.   Although the kinematical ages derived for the two shells have large uncertainties, they appear not to be coeval, the outer shell having nearly twice the kinematical age of the inner shell.  The exact evolution of the system is unclear, however the double-shelled structure is almost certainly a product of binary evolution, probably through two CFW or jet-like events.  Further investigation of the central binary is essential in order to understand its evolution and relate that to the exact formation mechanism at work in the nebula.

This study makes A~65 one of few to have had the link between nebular and binary inclinations examined, all of which have been found to be broadly consistent with theoretical predictions. As evidence mounts that binarity is responsible for the shaping in these PNe it is imperative to examine a larger sample, in order to begin to connect other morphological and kinematical parameters of the nebulae to those of their binary central stars, and therefore to the evolutionary processes at work in these systems \citep{jones2011,jones2012b}.  

\section*{Acknowledgments}

L.H. gratefully acknowledges the support of STFC and ESO through his studentships and thanks Wolfgang Steffen and Nico Koning for their help with {\sc shape}, and Marina Rejkuba for her helpful suggestions for the revision of this paper. This work was co-funded under the Marie Curie Actions of the European Commission (FP7-COFUND). We thank the staff at SPM and ESO Paranal for their support. We also thank the referee for their insightful comments, especially with regards the context of this work.

\bibliographystyle{mn2e}

\label{lastpage}

\end{document}